\documentclass[conference]{IEEEtran}
\IEEEoverridecommandlockouts
\usepackage{cite}
\usepackage{amsmath,amssymb,amsfonts}
\usepackage{algorithmic}
\usepackage{algorithm} 
\usepackage{graphicx}
\usepackage{textcomp}
\usepackage{xcolor}
\def\BibTeX{{\rm B\kern-.05em{\sc i\kern-.025em b}\kern-.08em
    T\kern-.1667em\lower.7ex\hbox{E}\kern-.125emX}}
    
\makeatletter
\newcommand{\linebreakand}{%
  \end{@IEEEauthorhalign}
  \hfill\mbox{}\par
  \mbox{}\hfill\begin{@IEEEauthorhalign}
}
\makeatother
\begin{document}
\title{A Scalable Blockchain-based Smart Contract Model for Decentralized Voltage Stability Using Sharding Technique }

\author{\IEEEauthorblockN{1\textsuperscript{st} Kimia Honari}
 \IEEEauthorblockA{ \textit{University of Alberta} \\ honari@ualberta.ca }
\and
\IEEEauthorblockN{2\textsuperscript{nd} Xiaotian Zhou}
 \IEEEauthorblockA{ \textit{University of Alberta} \\ xzhou9@ualberta.ca}
\and
\IEEEauthorblockN{3\textsuperscript{rd} Sara Rouhani}
 \IEEEauthorblockA{\textit{University of Manitoba}  \\sara.rouhani@umanitoba.ca }
\and
\IEEEauthorblockN{4\textsuperscript{th} Scott Dick}
 \IEEEauthorblockA{ \textit{University of Alberta}  \\ sdick@ualberta.ca }
\and
\IEEEauthorblockN{5\textsuperscript{th} Hao Liang}
 \IEEEauthorblockA{ \textit{University of Alberta} \\ hao2@ualberta.ca}
\linebreakand
\IEEEauthorblockN{6\textsuperscript{th} Yunwei Li}
 \IEEEauthorblockA{ \textit{University of Alberta}  \\ yunwei1@ualberta.ca }
\and
\IEEEauthorblockN{7\textsuperscript{th} James Miller}
 \IEEEauthorblockA{\textit{ University of Alberta}  \\ jimm@ualberta.ca }
}

\maketitle
\footnote{The paper is accepted as a regular paper in IEEE Blockchain Conference 2022}
\begin{abstract}
Blockchain technologies are one possible avenue for increasing the resilience of the Smart Grid, by decentralizing the monitoring and control of system-level objectives such as voltage stability protection. They furthermore offer benefits in data immutability and traceability, as blockchains are cryptographically secured. However, the performance of blockchain-based  systems in real-time grid monitoring and control has never been empirically tested. This study proposes implementing a decentralized voltage stability algorithm using blockchain-based smart contracts, as a testbed for evaluating the performance of blockchains in real-time control. We furthermore investigate sharding mechanisms as a means of improving the system's scalability with fixed computing resources. We implement our models as a proof-of-concept prototype system using Hyperledger Fabric as our blockchain platform, the Matpower library in MATLAB as our power system simulator, and Hyperledger Caliper as our performance evaluation tool. We found that sharding does indeed lead to a substantial improvement in system scalability for this domain, measured by both transaction success rates and transaction latency. 

\end{abstract}


\section{Introduction}
The ongoing development of the Smart Grid is putting increasing pressure on the monitoring and control systems that keep it stable. Renewable generation, for instance, is a growing part of our energy mix - but its intermittency demands rapid, real-time control actions to preserve voltage stability. Likewise, distributed generation and micro-generation further complicate the Smart Grid landscape. Meanwhile, the deployment of Intelligent Electronic Devices (lEDs) and Phasor Measurement Units (PMUs) - meant to enable fine-grained monitoring and control of the grid - has in turn created a tsunami of sensor data that threatens to overwhelm centralized control systems. Therefore, several studies have investigated distributed monitoring and control systems, in order to mitigate data volumes. This implies that the distributed control algorithms will principally operate on localized voltage measurements, but must still ensure system-level stability \cite{lee2021decentralized} \cite{mehrjerdi2012decentralized}\cite{mehrjerdi2013coordinated} \cite{schiffer2015voltage}. 

Blockchain technology and smart contracts are one possible approach to creating a decentralized IT infrastructure for the Smart Grid. In the past few years, there have been studies applying blockchains to various elements of the Smart Grid ecosystem. Some previous studies have  investigated  using smart contracts to provide ancillary services by tracking and managing energy distribution in the network \cite{di2019ancillary} or organizing some DERs to act as voltage regulators and curtail their individual power outputs \cite{danzi2017distributed}. A few studies also focused on scheduling and trading energy among Energy Storage Units (ESUs)  to minimize grid fluctuations \cite{baza2019blockchain} \cite{yang2019automated}. However, none of these studies have investigated or evaluated the capability and performance of blockchain as a real-time control system for complex objectives, using PMUs as the data sources. In the current paper, we design such a system for distributed voltage stability control. We have implemented an existing distributed voltage-stabilty control algorithm (DVS \cite{lee2021decentralized}) as a collection of smart contracts operating on a  permissioned blockchain. This is in contrast to the original implementation, which was built on the Distributed Coordination Blocks (DCBlocks)\cite{lee2016decentralized} framework, which is a classic distributed-computing platform.   

In addition, we propose a new transaction processing model based on the sharding technique to tackle the performance and scalability requirements of the Smart Grid. In this model, by growing the smart grid network, one or multiple local controllers can be assigned to different shards within the network, and a committee will be selected for each shard. This allows many more transactions to be processed in parallel at the same time, further improving the performance of the system.
We used Hyperledger Fabric \cite{androulaki2018hyperledger} to implement and evaluate our proposed model. For the Power flow simulation, we used the Matpower package \footnote{https://matpower.org/} \cite{zimmerman2016matpower} in Matlab. 
We analyzed the performance of our implementation using the Hyperledger Caliper benchmarking tools \cite{Caliper}, finding that our solution scales linearly with the number of shards compared to an un-sharded approach. To the best of our knowledge, this is the first study that presents a practical, scalable decentralized voltage stability algorithm based on blockchain technology. 

The remaining of this paper is organized as follows. In section II, we provide essential background in voltage stability and describe the DVS algorithm. In section III, we discuss the design of our smart contracts and the proposed model architecture on blockchain. Then we describe our second proposed model incorporating the sharding mechanism. In section IV, we describe our experimental methodology and results.  Finally, we offer a summary and discussion of future work in section V.

\section{Decentralized Voltage Stability (DVS) Algorithm}
In the electric power grid, power demand (load) must be less than or equal to the total available power. If the load in a given region exceeds the available power and no corrective actions are taken, the area voltages will become unstable and even collapse. This can severely damage electric systems, and so the response is to cut off the power flow to that region completely, resulting in a blackout. Voltage collapse or instability is a dynamic circumstance involving many nonlinear power system components, on a time scale that can range from seconds to hours \cite{van1998voltage}. 

Nowadays, ensuring voltage stability is a challenge due to the increasing number of renewable sources, increasing loads, environmental limitations on power system expansion, and high competition in the energy market. Voltage stability maintenance is an active process, which can be organized in a centralized or decentralized fashion. In the centralized voltage stability algorithms, a control center monitors the state of the power grid (or a subregion), commonly represented as one or more voltage stability indexes. These indexes essentially capture the grid's current safety margin against a voltage collapse. When stability is threatened (the indices approach dangerous values), the control center intervenes. Stored power from capacitors (purposefully charged for such needs) could be released, large-scale power customers might see large machines idled (such load-shedding contracts are common in energy markets), or additional generation may be brought on-stream. However, the volume and velocity of data received at the control center means that it may not react swiftly to instabilities in a smaller region. Decentralized algorithms, on the other hand, should be able to react much more swiftly to local instabilities. In this case, a voltage stability index would be calculated just for a limited area of responsibility, implying a far lower volume of data. The challenge, of course, is that a local algorithm will have far fewer resources available to address any instability. A mechanism for escalating interventions so as to call on resources beyond the local area will be needed \cite{lee2021decentralized}. 

Several studies have proposed or investigated different decentralized monitoring and control techniques, but none of them used blockchain and smart contracts \cite{rouhani2019security} as a distributed computing platform. Blockchain in this kind of system can increase trust, security, transparency, and the traceability of data shared across a power system network. We furthermore hypothesize that it provides a robust computational framework to implement local and distributed voltage stability algorithms. To test this hypothesis, we will implement one such algorithm as a set of blcokchain-based smart contracts. From the existing literature, we select the DVS algorithm \cite{lee2016decentralized}. First proposed in 2016, DVS includes both monitoring and control algorithms in their architecture. The authors validated the algorithm in simulations using the IEEE 30, 57, 118, and 300 bus topologies \cite{lee2021decentralized} (these are standardized power network simulations, with a varying number of connnections, used as research testbeds in power delivery systems research. Physically implementing and testing a new algorithm on the actual power grid is obviously inadvisable, and physical laboratory testbeds are either too simple to replicate actual power-system dynamics, or else prohibitively expensive.) The DVS algorithm is as follows:
\\
\noindent
\textbf{Initial Grouping and Group Formation:}

The DVS algorithm starts with an initial grouping method, which splits the power system grid and resources into multiple small groups. Each group is a set of nodes or substations aggregated based on their electric distance and network sensitivity. The substations or nodes in our blockchain-based model are able to instantiate smart contracts. They record and emit synchrophasor measurements from phasor measurement units (PMUs) to compute voltage stability indices \cite{gong2006synchrophasor} \cite{biswas2014synchrophasor}. They can take control actions based on those measurements, using local reactive power sources or Volt Var Compensators (VVCs). At a minimum, each group is assumed to have a few reactive power sources or VVCs available, as well as at least one transmission line and load and generator.

The buses in the local group can be classified into one of three categories: load bus, tie or boundary bus, and generation bus.  Load buses have a load connected to them. Generation buses have power sources or generators, and the tie buses are the interconnections among groups. For simplicity, they split each tie line in half and replace it with either a virtual load bus (virtual PQ bus) or a virtual generator (virtual PV bus) based on the power flow direction to represent power grid connections outside of the local group. 



\noindent
\textbf{DVS Monitoring Algorithm:}

The DVS monitoring algorithm estimates a voltage stability index (VSI) using the proposed Thevenin's Equivalent approach in \cite{gong2006synchrophasor}. The equivalent voltage and impedance are calculated based on each group's information for representing the external system connected to each load bus. First, they create an admittance matrix for each group for representing the network topology. Equation (\ref{eq2}) shows the admittance matrix, in which the G denotes generation bus, L represents Load bus, and T shows the tie bus. Therefore, \(Y_{GL}\), \(Y_{GT}\), \(Y_{TL}\), \(Y_{TT}\) , \(Y_{TG}\), and \(Y_{LL}\) are the admittances between generator to load, generator to tie line, tie line to load, tie line to tie line, tie line to generator, and load to load, respectively. 

\begin{equation} \label{eq2}
\begin{split}
Y=
  \begin{bmatrix}
Y_{GL} & Y_{GT} & Y_{GG}\\
Y_{TL} & Y_{TT} & Y_{TG}\\
Y_{LL} & Y_{LT} & Y_{LG}
\end{bmatrix}
\end{split}
\end{equation}

\begin{equation} \label{eq3}
\begin{split}
Z_{th}=Z_{LL}= (Y_{LL}-Y_{LT}Y_{TT}^{-1}Y_{TL})^{-1}
\end{split}
\end{equation}

\begin{equation} \label{eq4}
\begin{split}
V_{th_j}=((\frac{S_{L_j}}{V_{L_j}})^{*} * Z_{th_j} ) - V_{L_j}\\ j=1,...,n\ n=number\ of\ load\ bus. \\
S_{L_j} = complex\ power\ flow\ out\ of\ bus\ j.\\
V_{L_j}= voltage\ magnitude\ of\ load\ bus\ j.
\end{split}
\end{equation}

Thevenin's parameters are calculated using equations (\ref{eq3}) and (\ref{eq4}). The \(V_{th}\) (voltage equivalent) and \(Z_{th}\) (equivalent impedance) are then used to approximate maximum active power (\(P_{max}\)), maximum reactive power (\(Q_{max}\)), and maximum complex power (\(S_{max}\)) for each load bus \cite{gong2006synchrophasor}. (Recall that alternating current power follows a sine-wave pattern, which is analyzed using complex values; reactive power is the imaginary component of the complex power vector.) Similary, (\(P_{load}, Q_{load}, S_{load}\)) are the real, reactive, and complex power values of load at each bus, respectively (please refer to \cite{gong2006synchrophasor} for a detailed derivation). Using calculated maximum power for each bus, the VSI for each load bus is calculated as follows:

\begin{equation} \label{eq5}
\begin{split}
VSI=Min(\frac{P_{max} -P_{load}}{P_{max}},\frac{Q_{max} -Q_{load}}{Q_{max}}, \frac{S_{max} -S_{load}}{S_{max}}).
\end{split}
\end{equation}

The VSI for each bus is used in the DVS algorithm to detect buses with a weak stability margin; the DVS control algorithm is then used to prevent voltage collapse.

\noindent
\textbf{DVS Control Algorithm:}

When a bus is found to have a weak stability margin, additional reactive power from the closest source (in terms of electrical distance) is injected into the network. 
 In this algorithm, a Priority Index (PI) matrix is first formed for each group using the admittance matrix. The PI matrix represents an electrical distance between buses and helps to find the closest VVCs to the weak bus. Based on this matrix, the top priority is given to a VVC directly connected to the weak bus; and the next set of priorities is given to reactive power sources based on the ascending ranking of electrical distance. After selecting the closest reactive power, the required reactive power to adequately raise the stability margin at the weak bus is calculated using the Jacobian Matrix. In Equation (\ref{eq6}), \(V_{req} \) is the minimum acceptable voltage, \(Q_{req} \)is the required reactive power, and \(V_{W eakBus}\) is the voltage magnitude of the weak bus.



\begin{equation} \label{eq6}
\begin{split}
Q_{req}=\frac{\delta Q}{\delta V} * (V_{req})-V_{WeakBus}).
\end{split}
\end{equation}

\noindent
The \(\frac{\delta Q}{\delta V}\) term represents the sensitivity of the bus where the VVC is located versus the weak bus. In the DVS algorithm, the chain rule is used to calculate the sensitivity of buses that are not directly connected to the weak bus. If needed, the control algorithm may run multiple times to compensate for the voltage stability at the target bus. If the DVS algorithm cannot find enough reactive power sources within the group to correct the stability margin, it then merges the group with an adjacent group, and calls on that group's resources as well. This process will continue, drawing in resources from a wider and wider area until enough additional power is injected to restore stability.

\section{Blockchain-Based Smart Contract Design for DVS Algorithm}

Blockchain technology offers security, traceability, decentralization, and immutability; characterisitcs which can leveraged in implmenting the DVS algorithm. The blockchain model provides a secure platform for  each group to communicate with each other and reach a consensus on what buses require control action, and how best to organize the same. Utilizing smart contracts enables us to automate these procedures, thus reacting faster to voltage instabilities; as noted previously, a voltage collapse can potentially happen in just seconds. At the same time, the security and immutability of the blockchain offer enhanced protection and auditability for critical infrastructure. While other authors have explored blockchain systems in the past, they either do not use blockchain as a platform for decentralized voltage stability, or neglect the performance of their consensus mechanism if they do \cite{lee2016decentralized}. 

This study proposes and evaluates a blockchain-based smart contract model to implement the DVS algorithm. We furthermore address scalability; plainly, in a large-scale grid there will be many groups, and transactions between groups. The blockchain network underlying the grid could thus be overloaded. Therefore, we designed a sharding solution for our blockchain-based DVS algorithm. We formally discuss the design of our model for the sharding mechanism based on two consensus levels: shard-level consensus and mainchain consensus. This section will elaborate on our blockchain-based approach and workflow and then describe our sharding mechanism to improve scalability.

\subsection{Workflow}

\begin{figure}
    \centering
    \includegraphics[width=\columnwidth]{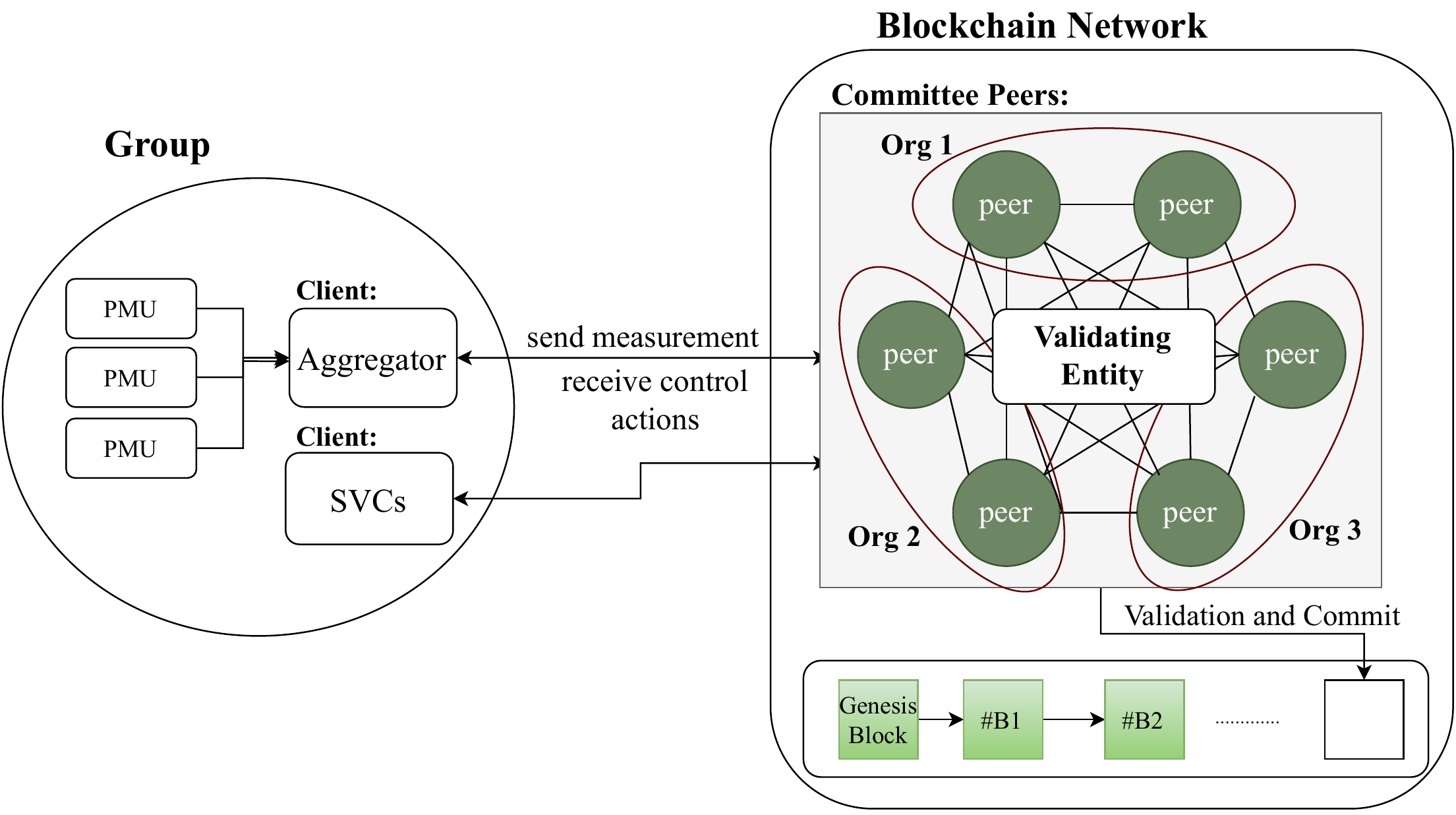}
    \caption{Blockchain based architecture for the DVS algorithm - without sharding mechanism. }
    \label{fig:noshard}
\end{figure}

In this subsection, we discuss the entire workflow of our proposed framework and elaborate on the network topology and smart contract details for DVS monitoring and control algorithms.  Firstly, we categorize the participants in our network, following the notation in Fig. \ref{fig:noshard}.

\textbf{Clients:}

We assume that multiple PMU devices are distributed inside each group for measuring phasor quantity. PMU devices estimate the magnitude and phase angle of an electrical phasor quantity such as voltage or current in the power grid. PMUs can report high temporal resolution measurements, up to 120 Hz \cite{wilson1994pmus}. Several papers investigated the optimal location of PMUs when the number of PMUs is limited \cite{abbasy2009unified}\cite{carrion2018optimal}, which is out of the scope of this paper.

In order to gather measurements, we assume one or multiple computing devices are responsible for aggregating the PMUs measurement in each group and instantiating relative smart contracts through the Fabric SDK. The Hyperledger Fabric SDK allows applications to interact with a Fabric blockchain network via a simple API to submit transactions to the ledger or query the recorded data with minimal code. Furthermore, the reactive power sources and VVCs are also connected to the network and update their available resources on the ledger for control actions. In the case of need, smart contracts can send a control action to activate or deactivate VVCs on the grid. For simulating the client-side, we use the Matpower library in Matlab to run optimal power flow and inject reactive power to each bus. For interacting with Fabric SDK, the related data of each group is converted to JSON. Then the data is sent via an HTTPS  request to the RESTful Web-service to submit or query a transaction on the ledger.
 
\textbf{Organization and Peers:}

 The HyperLedger Fabric blockchain network is built up from the peers owned and contributed by the different organizations. Each organization can have one or multiple peers responsible for various tasks and offering API services for clients. The smart contracts are executed on the peers, and peers maintain copies of the ledger. We assigned one or multiple peers to each group.

\textbf{Committee Peers:}

Members of the committee are responsible for evaluating and validating every task and data during consensus. In a network with no sharding mechanism, we assume all peers in the blockchain network are responsible for validating and endorsing every executed task.

\vskip 0.2in

The Clients can instantiate and submit different transactions on the network, as managed by the smart contracts. We have implemented four primary tasks of the DVS algorithm as smart contracts:

\textbf{Initial grouping:} In our prototype system, we assume that all admittance values are constant, and the system's topology will not change; therefore, this transaction will be executed once during the initializing of the network. The admittance matrix, Priority Index matrix, and the constant part of the Jacobian matrix are calculated through Matlab libraries, and their data are recorded for each group and combination of groups on the ledger. Note, however, that these values would change if the topology of the network changed.

\textbf{ComputeVSI:} This transaction receives the data of PMU measurements and uses the DVS algorithm to calculate the VSI for each load bus. Then the VSI values are sorted via the shell sort algorithm, and if the minimum VSI is less than the specified threshold, the LocalController transaction will be called. This transaction also reads  information from the ledger, such as the impedance values of the local group determined during the initial grouping. Algorithm \ref{computeVSI} shows the pseudo-code of ComputeVSI.

\begin{algorithm}
	\caption{ComputeVSI} 
	\begin{algorithmic}[1]\label{computeVSI}
	    \STATE \textbf{Input:} PMU measurements, GroupId
		\STATE Impedance= Retrieve the data from the ledger(GroupId)
		\STATE Calculate \(V_{th}\) and \(Z_{th}\) for each load bus (3)(4)
		\STATE Calculate VSI for each load bus (5)
		\STATE Sort VSI ascending
		\IF {$VSI[0]\leq $Threshold } 
           \STATE 
            Call LocalController
        \ELSE
           \IF{ mergedgroup == true}
             \STATE Split the group
            \ENDIF
        \ENDIF
	\end{algorithmic} 
\end{algorithm}

\begin{algorithm}
	\caption{LocalController} 
	\begin{algorithmic}[1]\label{LocalController}
		\STATE \textbf{Input:} WeakBus, Resources, GroupId
		\STATE PI, Jacobian= Retrieve  the data from the ledger(GroupId)
		\STATE w=WeakBus
		\STATE Calculate \(\frac{\delta Q_{w}}{\delta V_{w}}\)
		\STATE Calculate Q\_req (6)
		\IF{ resource available at WeakBus}
		    \STATE Activate a VVC at WeakBus
		\ELSE
		  \STATE List= Create PI list 
		  \WHILE{List is not Empty}
		  \STATE i=List.GetNextPriority()
		  \STATE Calculate \(\frac{\delta Q_{i}}{\delta V_{w}}\)
		  \STATE Calculate Q\_req (6)
		  	\IF{ Resource available at Bus i}
		    \STATE Activate a VVC at Bus i
		    \STATE Break
		    \ENDIF
		  \ENDWHILE
		\ENDIF
	 	\STATE Save VVC\_index, Q\_req,and GroupId on the ledger
	 
	\end{algorithmic} 
\end{algorithm}

\textbf{LocalController:} This transaction will be called inside of ComputeVSI to calculate the required reactive power to stabilize a weak bus, and activate VVCs to deliver it. The steps of this transaction are shown in Algorithm \ref{LocalController}. It first retrieves the PI and Jacobian matrix from the ledger. Then it finds the electrically closest VVC to the weak bus and calculates the required reactive power. Two libraries (Jacobian, and List) were developed to help implement the DVS algorithm on smart contracts. The first one is used to calculate Jacobian Values with real-time voltage angle and magnitude, and the second one is used to find the top priority buses using the PI matrix. In the end, we save the amount of reactive power that is injected into the bus on the ledger. This data can be used to calculate the amount of money that we owe to each source, or predict the amount of power we may need in the grid in different periods, allowing us to actively prevent voltage collapse.

\textbf{GlobalController:} Suppose the LocalController transaction cannot find enough power resources. In that case, the system will call this transaction to find an adjacent group with additional available resources, and send control actions to merge the data of these two groups. After that, the ComputeVSI and LocalController will be called on the merged data until the weak bus is stabilized. Once this is accomplished, the two groups will again be split. In merging the groups, we consider that one of the group aggregator devices is responsible for aggregating all PMUs of two groups.

\subsection{Sharding Mechanism}

Adding sharding to the blockchain-based DVS allows scalability by splitting the blockchain network into smaller subgroups. The DVS is a  real-time control algorithm and needs to react to each change in the power grid to prevent instability. However, as the size of the power grid controlled by DVS grows, it becomes likely that the many components of the grid are in constant flux. Hence, transaction volume and latency can be serious issues, which may delay and thus undermine DVS control actions in a rapidly-evolving voltage collapse. The sharding mechanism is designed to reduce storage and communication requirements while increasing throughput as the number of shards increases (while the physical network size remains constant) \cite{xie2019survey}. Sharding techniques and the scalability of blockchains have been investigated in various studies and applicaitons such as IoT networks \cite{tong2019hierarchical}, federated learning \cite{madill2022scalesfl}\cite{yuan2021chainsfl}, and 6G networks \cite{lu2020low}. Some of the studies also proposed a new blockchain protocol and cross-shard techniques such as RapidChain \cite{zamani2018rapidchain}, and OmniLedger \cite{kokoris2018omniledger}. 

\begin{figure}
    \centering
    \includegraphics[width=\columnwidth]{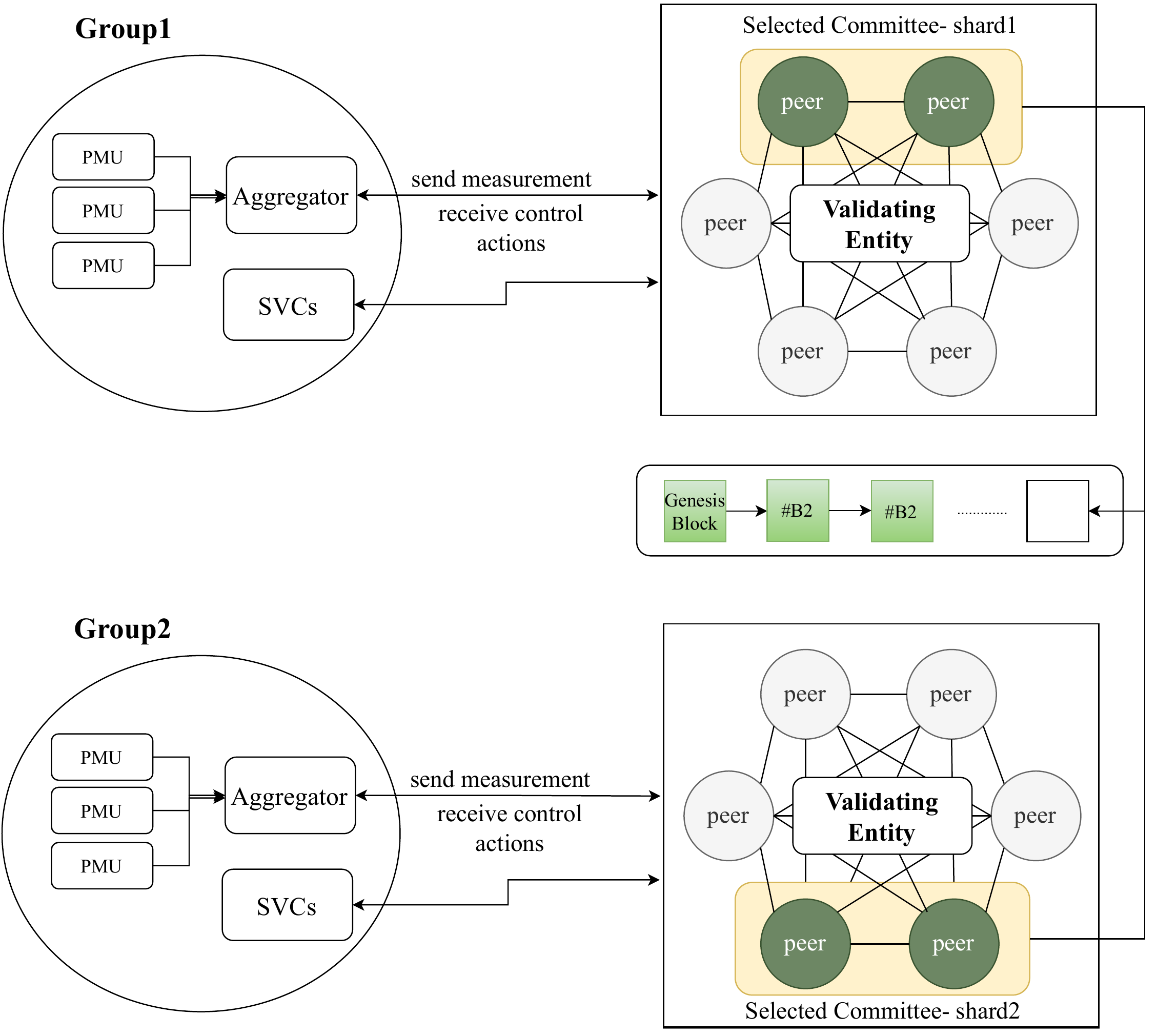}
    \caption{Blockchain based model architecture for the DVS algorithm - sharding mechanism. }
    \label{fig:shard}
\end{figure}

In this paper, we employed a sharding technique to make our system more scalable and adoptable by expanding the network. In this technique, instead of using all peers to endorse and validate tasks in the system, we assign a smaller group of committee peers for each shard. Each shard can be responsible for validating shard level transactions of one or multiple adjacent groups. Fig. \ref{fig:shard} provides an overview of our sharding model. 

 If we assign enough reactive sources for each load bus in a group, the chance of solving the voltage stability problem in each group will be increased. Therefore, we can split transactions based on the importance and occurrence frequency to shard level and mainchain level. Shard level refers to the transactions executed and endorsed by a subset of peers in each shard, and mainchain level refers to the transactions executed by all peers in the network. At the mainchain level, we assume that the global controller and all data related to the topology of each group will be executed and validated by all peers or a subset of peers assigned to every single group. This will help to preserve high security for critical control actions and access to the data of all groups without needing cross-shard communication. The other transactions such as ComputeVSI and LocalController, which occur within a single group, are executed in the shard level consensus. 

\begin{table*}[!t]
\centering
\caption{Experimental Configuration}
\label{Table.exp}       
\begin{tabular}{cccccc}
\hline
Component & Version & CPU & GPU & RAM & Disk (SSD)
\\
\hline
Caliper Benchmark & Caliper 0.4.2 & Intel Core i7-9700K & GeForce RTX 2080 TI & 62.8 GB & 500 GB (SSD) \\
\hline
Fabric peer & Fabric 2.3.3 & Intel Core i7-9700K & GeForce RTX 2080 TI & 62.8 GB & 500 GB (SSD) \\\hline
Matpower &  Matpower 7.1 & Intel Core i7-9700K & GeForce RTX 2080 TI & 62.8 GB & 500 GB (SSD) \\
\hline
\end{tabular}

\end{table*}

\section{Experiment}

\begin{figure}
    \centering
    \includegraphics[width=\columnwidth]{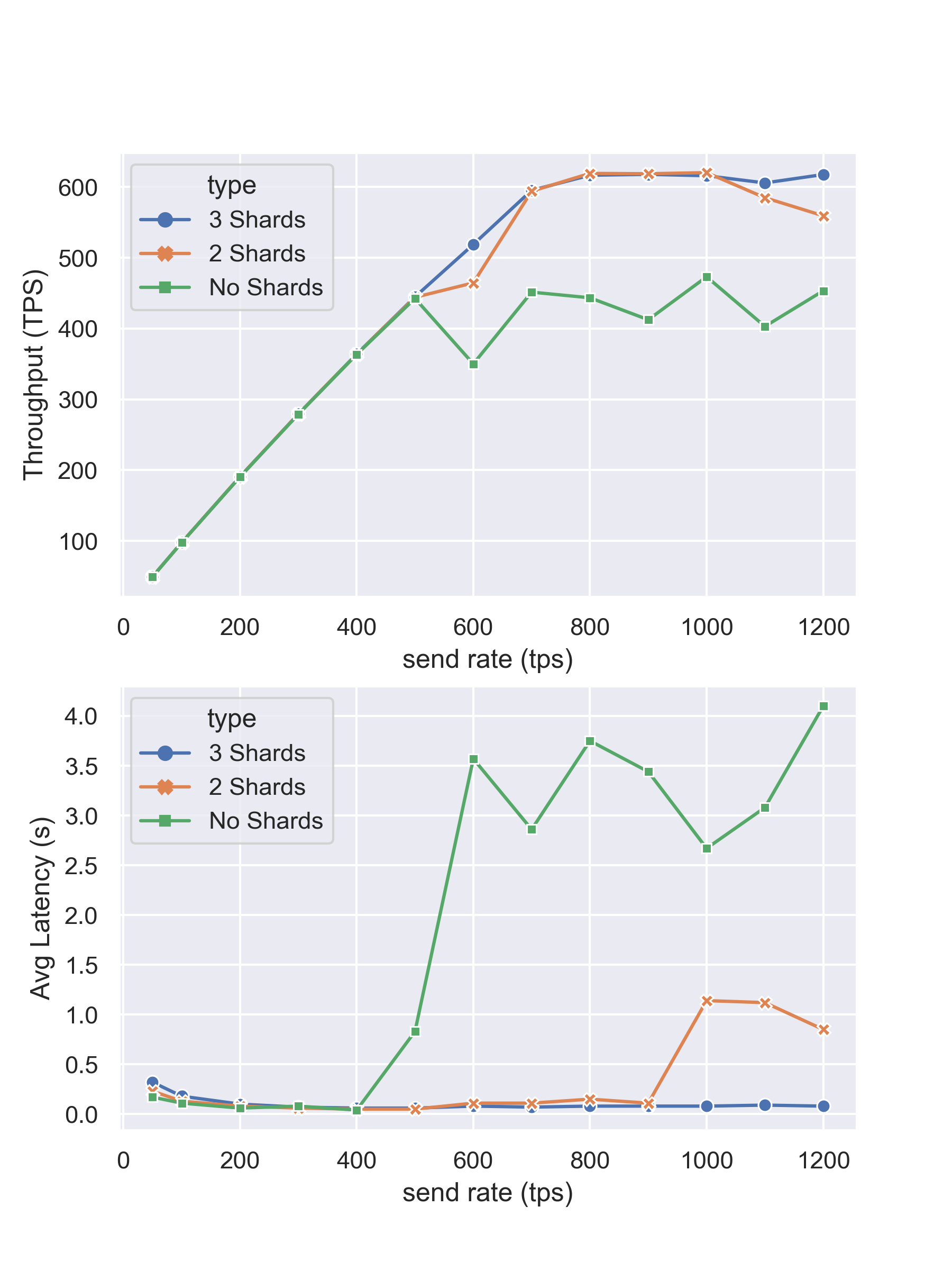}
    \caption{Send rate vs. system throughput (TPS) \& Average
response latency - Fixed values: 3 worker over 8000 transactions - ComputeVSI transaction. }
    \label{fig:result1}
\end{figure}

\begin{figure}
    \centering
    \includegraphics[width=\columnwidth]{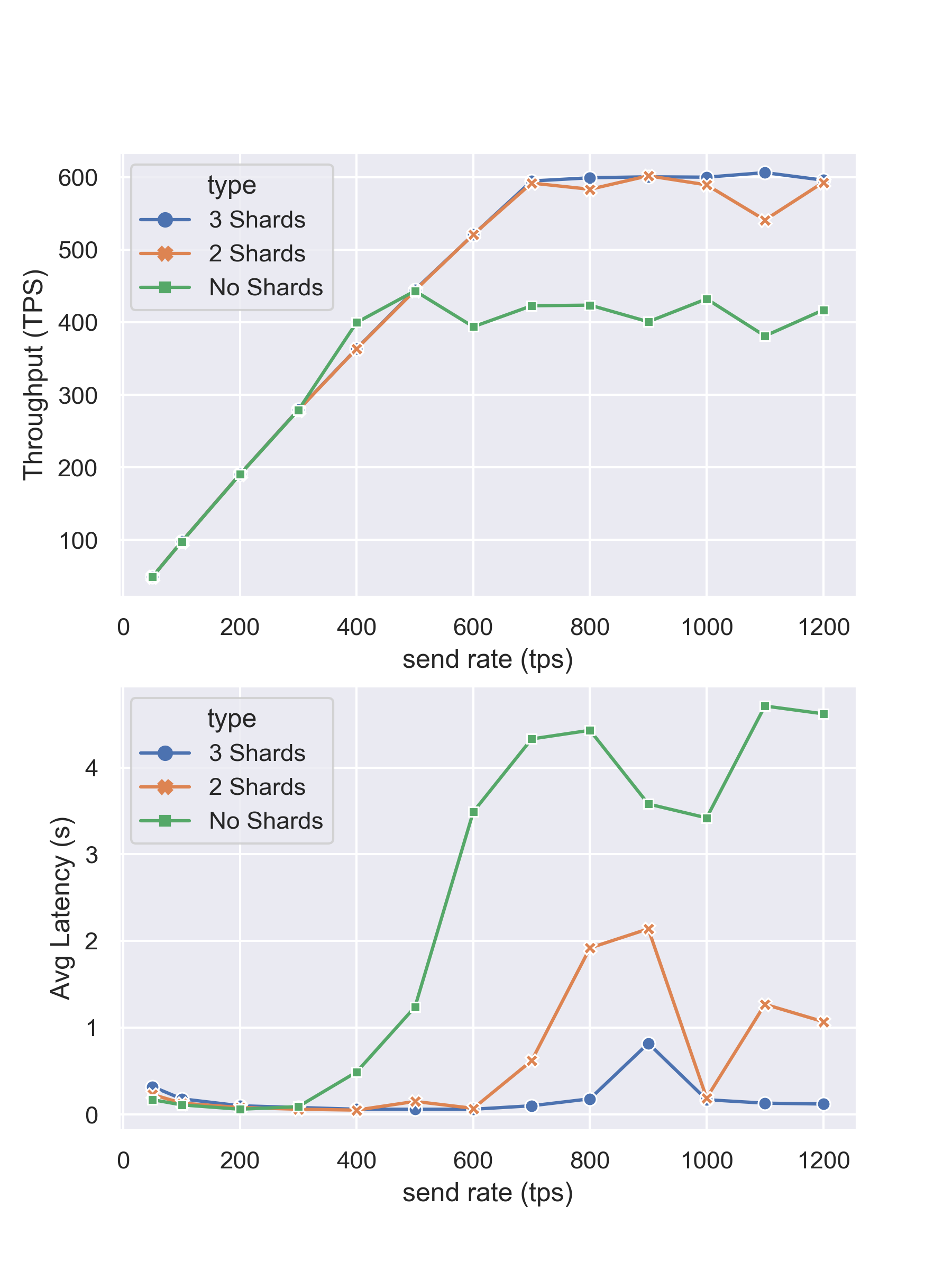}
    \caption{Send rate  vs. system throughput (TPS) \& Average
response latency - Fixed values: 3 worker over 8000 transactions - ComputeVSI+LocalController transactions. }
    \label{fig:result2}
\end{figure}

For our experiments, we seek to quantify what benefit, if any, is provided to our prototype system by implementing sharding. Accordingly, we first measure throughput and latency without sharding, and then compare this with specific levels of sharding; specifically, a model with two shards, and a model with three shards.

We used Hyperledger Fabric \footnote{https://hyperledger-fabric.readthedocs.io/en/release-2.2/} \cite{androulaki2018hyperledger}, a permissioned platform, to implement these three models \footnote{https://github.com/Scalable-Blockchain-Systems/DVSCode}. The Fabric has a modular and configurable architecture and supports plug-n-play consensus and membership services. The execute-order-validate architecture for transactions provided by Fabric allows each peer to evaluate models in parallel within each shard, as opposed to most public blockchain platforms that have first-order transactions and execute them sequentially. Fabric also provides communication through channels that is a private layer of communication between two or more specific network members such as organizations and peers. Each transaction executed on a channel must be authenticated and authorized by channel members to transact on that channel. Peers can be members of multiple channels and perform channel related operations.

Networks run locally on a single machine simulating a Fabric test-network with a single orderer running Raft \cite{ongaro2014search}. For the no-shard model, we consider six peers, each owned by different organizations and a certificate authority. Every group sends their requests to three different peers, and all peers are responsible for validating and endorsing a transaction. In the 2-shards model, we consider that each shard has three peers that evaluate models in parallel within each shard. For the 3-shards model, we consider each shard has two peers responsible for validating each group's requests. We implement two smart contracts, one for ComputeVSI and local transactions called VSIContract, and the other for handling Global Controler called GlobalContract. The smart contract is known as a chaincode in Fabric and deployed to a specific channel. The channels are used to simulate shards in the system, where each channel operates independently with the ability to have different membership and endorsement policies. All six peers are a member of the "mainchain" channel that we deployed the GlobalContract smart contract on, and we deployed a VSIContract to each shard channel. 

For the power grid simulation, we applied the DVS  algorithm to the IEEE 30-bus network   \cite{alsac1974optimal} (also used in \cite{lee2021decentralized}). Following \cite{lee2021decentralized}, we split the grid into three different groups and save the admittance, PI matrix, and constant part of Jacobian for each group on the ledger. We begin with a normal scenario, with no voltage stability problems. We collected this normal data for testing a ComputeVSI transaction with Hyperledger Caliper. Then, we increase the load of some of the buses, decreasing the stability margin  to trigger the LocalController transaction.

For evaluation, we used Hyperledger Caliper \cite{Caliper} which is an open-source benchmarking tool allowing simulation of various workloads for our system. An independent process is responsible for sending a transaction with a specified configuration, and the system's performance (latency, throughput, success rate) is reported. We conducted tests with varying numbers of workers, transactions numbers, and send rates (TPS)  to evaluate the limits of each model for ComputeVSI and LocalController transactions. The experimental configuration is summarized in
Table \ref{Table.exp}.

\begin{figure}
    \centering
    \includegraphics[width=\columnwidth]{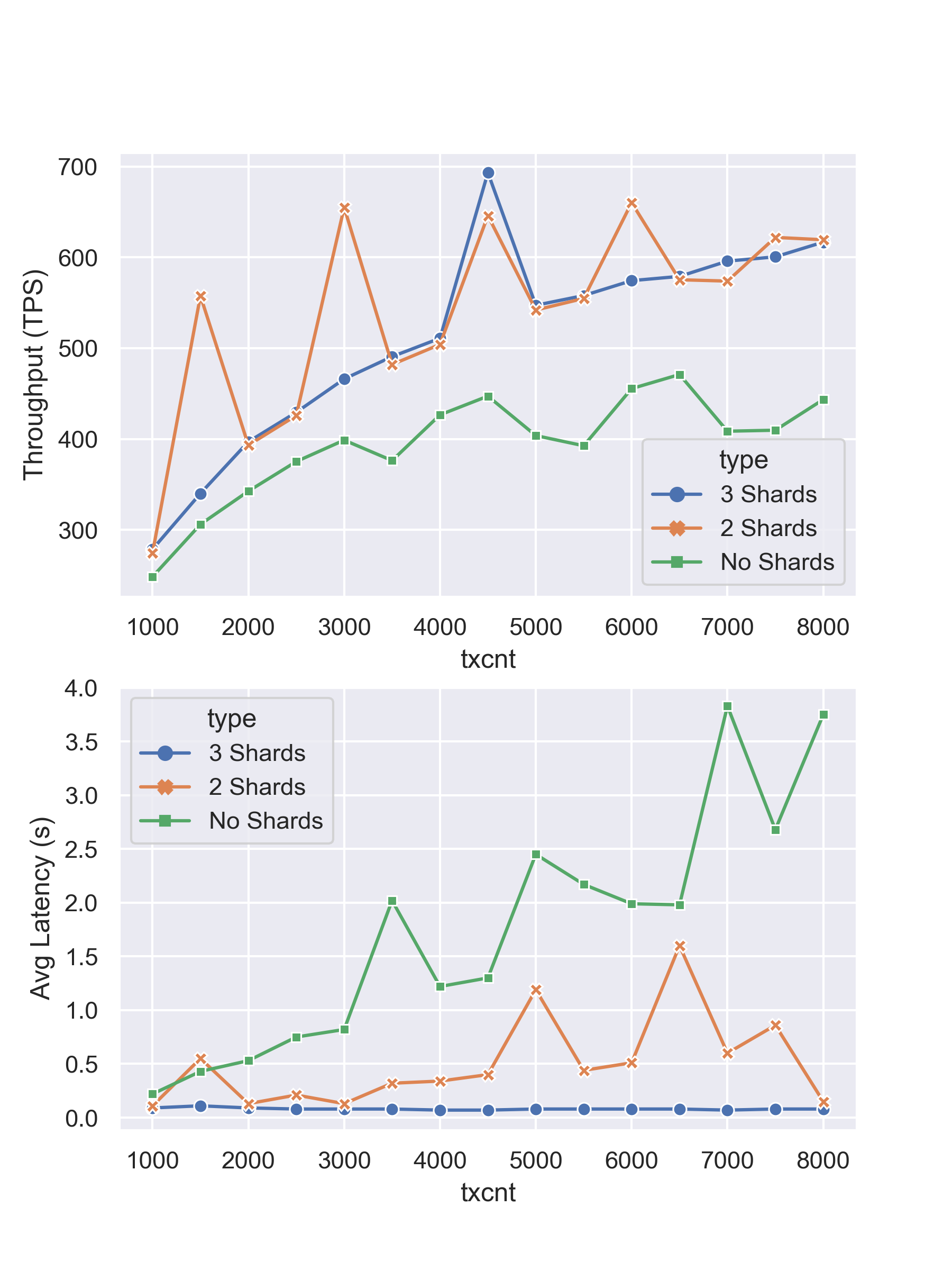}
    \caption{Transaction number (txcnt) vs. System throughput (TPS) \& Average
response latency - Fixed values: 3 worker with tps of 800 - ComputeVSI transaction. }
    \label{fig:result3}
\end{figure}

\begin{figure}
    \centering
    \includegraphics[width=\columnwidth]{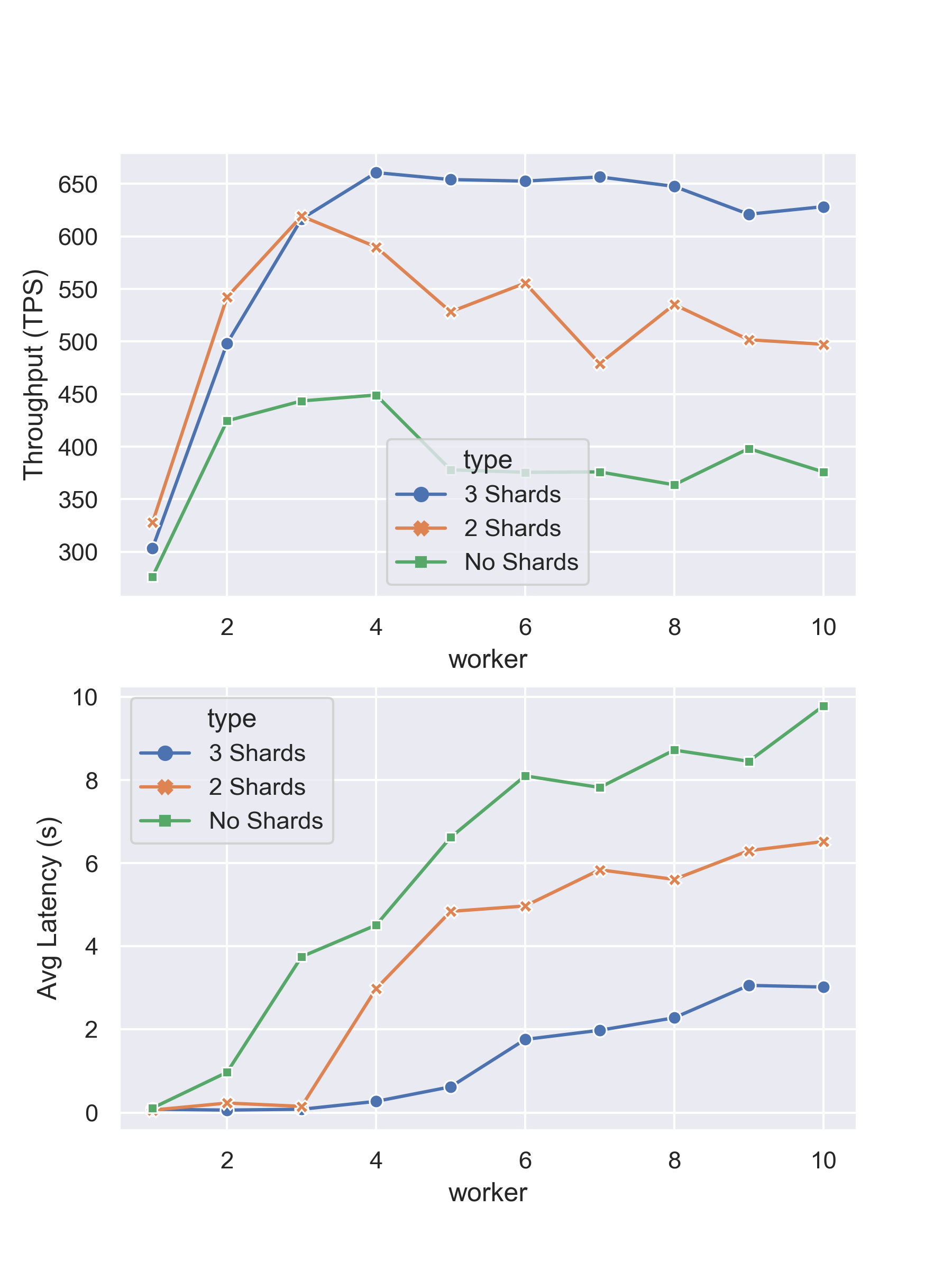}
    \caption{Number of workers  vs. System throughput (TPS) \& Average
response latency - Fixed values: 8000 transaction with tps of 800 - ComputeVSI transaction. }
    \label{fig:result4}
\end{figure}

\subsection{Result}

The ComputeVSI transaction is the only transaction that will be called multiple times every second (based on the measurement frequency of the PMUs). In contrast, the Control transaction will be called a few times when needed. Therefore, most of the workload is run with the normal data, invoking just the ComputeVSI transaction. 

To test the maximum throughput achieved by our system, we measure the send rate against the system throughput and average latency one time with normal data and the other time with unstable data. Fig. \ref{fig:result1} shows the result for normal data that just run the ComputeVSI transaction, and Fig. \ref{fig:result2} shows the result for unstable data that run both ComputeVSI and LocalController transactions. These workloads are run with 3 Caliper workers over 8000 transactions. By increasing the send rate, each model will reach a point that becomes saturated and is unable to handle a higher send rate. As shown in the figures, the no-shard model gets to this threshold sooner than the sharding model.

To explore the limits of a usage surge, we tested the number of transactions sent by the system with respect to throughput and average latency.  This workload is run with 3 caliper workers and fix sent rate (tps) of 800. Fig. \ref{fig:result3} shows the throughput and latency of this workload. As can be seen, the result shows that the sharding model can significantly improve the overall throughput and latency. 

Finally, we tested how the system handles concurrent requests by running multiple workloads, varying the number of caliper workers. This workload configuration sends 8000 transactions with a send rate of 800 to measure the system throughput and average latency. The number of caliper workers allows us to scale the workload generation, and each worker processes performs the actual workload generation in parallel and independently of each other. As we can see in  Fig. \ref{fig:result4},  the throughput of the system has a general downward trend in the system throughput with respect to the number of workers. Similarly, we can see an upward trend in average latency. We can see that the number of shards plays the most important role for average latency due to these workloads being able to operate in parallel across shards.

\section{Conclusion and Future Works}
We designed and implemented a scalable blockchain-based DVS system to compute a voltage stability index and take control actions in the electrical power system. Our smart contracts regulate automatic computations of decentralized DVS algorithm and manage resources in the electric power system.


The experimental results demonstrate that permissioned blockchains can handle a large number of power grid transactions in a few seconds, and we can improve system performance close to linearly with the addition of shards. This helps address the scalability issue related to blockchain consensus in a large-scale power grid network.

In future work, we will analyze the trace and pattern of data associated with voltage collapse to identify potential threats in the system, predict possible vulnerabilities in the power system, and integrate automatic preventative action. This will also provide data on the required number of reactive power sources that need to be added to improve the grid's stability. Hence, we will investigate the Machine learning models and AI techniques integrated with our model to reduce the complexity or give better intuition about the system.

\bibliographystyle{IEEEtran}
\bibliography{conference_101719}

\end{document}